\documentclass[aps,preprint]{revtex4}
\usepackage{graphicx,color}

\begin{document}
\title{Reflexion and Diffraction of Internal Waves analyzed with the
Hilbert Transform}

\author{Matthieu J. Mercier, Nicolas B. Garnier and Thierry Dauxois} \date{\today}
\affiliation{Universit\'e de Lyon, Laboratoire de Physique de
l'\'Ecole Normale Sup\'erieure de Lyon, CNRS, 46 all\'{e}e d'Italie,
69364 LYON cedex 07, France}
\bibliographystyle{plain}

\date{\today}

\begin{abstract}
We apply the Hilbert transform to the physics of internal waves in
two-dimensional fluids. Using this demodulation technique, we can
discriminate internal waves propagating in different directions:
this is very helpful in answering several fundamental questions in
the context of internal waves. We focus more precisely in this paper
on phenomena associated with dissipation, diffraction and reflection
of internal waves.

\end{abstract}
\maketitle

\section{Introduction}

The Synthetic Schlieren technique \cite{Dalziel2000} is a very
powerful method to get precise and quantitative measurements for
two-dimensional internal waves in stratified fluids. Such a
technique was very effectively used to get quantitative insights
while studying different mechanisms for internal waves. Let us just
mention the emission, propagation and reflection of internal
waves~\cite{Sutherland1999,PeacockWeidman,PeacockTabei,Gostiaux_these,NyeDalziel},
or the generation and reflection of internal
tides~\cite{Gostiaux_these,GostiauxDauxois2007,Peacock2008}.
However, when considering internal waves generated by an oscillating
body or by an oscillating flow over a topography, the analysis is
drastically complicated by the possibility of different directions
of propagation associated to a single frequency; such a problem
arises also when multiple reflections occur at boundaries.


We present in this article a method to discriminate the different
possible internal waves associated with one given
frequency~$\omega$. These waves can be discriminated by their wave
vectors $\mathbf{k}=(k_{x},k_{z})$, according to the sign of both
components, $k_{x}$ and $k_{z}$. The transformation we present here
not only offers an analytical representation of the wave field which
allows us to extract the envelope and the phase of the waves, but
allows also to isolate a single wave beam. This method is based on
the Hilbert transform (HT) previously applied to problems dealing
with propagating waves, but adapted here to two-dimensional
phenomena.

The method is used here to tackle several fundamental issues in
order to bring new insights. It is important to emphasize that we
used a source of monochromatic internal plane waves to facilitate
the comparison with theoretical results.

The paper is organized as follows. In
section~\ref{hilberttransform}, we present the Hilbert transform. In
the  section~\ref{Applications}, we present its application to the
classical oscillating cylinder experiment, with a special emphasis
on the insights provided by the Hilbert Transform.  In
section~\ref{newapplications},  we study three different physical
situations that can be nicely solved with this technique. The
dissipation length is studied in Sec.~\ref{dissipation}, the back
reflection on a slope in Sec.~\ref{backreflection}, while
Sec.~\ref{diffraction} focus on the diffraction mechanism. Finally
section~\ref{conclusion} concludes the paper.

\section{Principle of the Hilbert Transform}
\label{hilberttransform}

\subsection{Presentation of the variables}

Before explaining on a simple example the different steps necessary
to apply the Hilbert transform, let us briefly recall different
properties of internal gravity waves. We consider a two-dimensional
$(x,z)$ experimental situation and denote $t$ the time variable, $g$
the gravity and $\rho(x,z)$ the density. In a linearly stratified
fluid such that ${\partial\rho}/{\partial z}<0$, and within the
linear approximation, it is well-known~\cite{K90} that the same wave
equation
\begin{equation}\label{ODE_classic}
\triangle \psi_{tt} + N^2\psi_{xx} = 0
\end{equation}
is valid for the field $\psi(x,z,t)$ which stands for either the
streamfunction, both velocity components, the pressure, or the
density gradients. The constant
\begin{equation}N=\sqrt{-\frac{g}{\rho}\frac{\partial\rho}{\partial
z}}\end{equation} which characterizes the oscillation of a fluid
particle within a linearly stratified fluid is the so-called
Brunt-V\"{a}is\"{a}l\"{a} frequency. Looking for propagating plane
waves solutions
\begin{equation}\displaystyle \psi=\psi_0\, e^{i(\omega t -
\mathbf{k.x})}\,,\end{equation} where $\mathbf{x}=(x,z)$ and
$\mathbf{k}=(k_x,k_z)$, one gets the dispersion relation
\begin{equation}
\omega^2 = N^2 \frac{k_{x}^{2}}{k_{x}^{2} + k_{z}^{2}} = N^2
\sin^2\theta\,,\label{dispersionrelation}
\end{equation}
if one introduces $\theta$ the angle between the wavevector and the
gravity. In this unusual dispersion relation, it is apparent that
changing the sign of the frequency $\omega$ or the sign of any
component $k_x, k_z$ of the wavevector has no consequences. So four
possible wavevectors are allowed for any given positive frequency,
smaller than the Brunt-V\"{a}is\"{a}l\"{a} frequency.

The Synthetic Schlieren technique gives quantitative measurements of
the respectively horizontal and vertical density gradients,
$\rho_x(x,z,t)$ and $\rho_z(x,z,t)$. As anticipated, both quantities
verify Eq.~(\ref{ODE_classic}). In the remainder of this section, we
work on a field $U(x,z,t)$, which might be either $\rho_x(x,z,t)$,
$\rho_z(x,z,t)$, or a velocity component as obtained in PIV
experiments.

\subsection{A simple one-dimensional example}

We present here, in the first stage, how to compute the
complex-valued field $\tilde{U}(x,z,t)$ such that $U(x,z,t)$ will
correspond to its real part $Re(\tilde{U}(x,z,t))$. In order to do
so, we demodulate the signal by applying the Hilbert transform. To
avoid misunderstandings, let's note that the Hilbert transform is
sometimes the name of the operation that associates the real-valued
field $Im(\tilde{U})$ to the real-valued field $U$, such that the
complex number $\tilde{U}$ can be fully reconstructed. In this
article, we call Hilbert transform (or complex demodulation) the
operation associating the  complex-valued field $\tilde{U}$ to the
real field~$U$. This demodulation technique has been previously used
to compute local and instantaneous amplitudes, frequencies and
wavenumbers~\cite{Croquette:89:2, Garnier:03:1, Garnier:03:2} but,
to the best of our knowledge, the present study is the first
application in the context of internal gravity waves.

As an introductory example, let us consider a simple signal in one
spatial dimension constructed as the superposition of two wave beams
propagating in the vertical direction $z$ with the same frequency
and the same wavenumber $k_z$
\begin{equation}
U(z,t) = A \cos(\omega t - k_z z) + B \cos(\omega t + k_z z).
\end{equation}
As $z$ is the vertical component, the first term corresponds to a
wave propagating upward, whereas the second one to a wave
propagating downward. For the sake of simplicity, we further suppose
that the amplitude $A$ and $B$ are constant in space and time.
Rewriting the cosines as the sum of exponentials with complex
arguments, and decomposing according to a Fourier transform in time,
we have
\begin{equation}
U(z,t) = \hat{U}_1\, e^{i \omega t} + \hat{U}_2\, e^{-i \omega t}
\end{equation}
where $\hat{U}_1 = (Ae^{-i k_z z} + Be^{i k_z z})/2$ and $\hat{U}_2$
is the complex conjugate of $\hat{U}_1$. So if we filter out the
negative frequencies in Fourier space, and multiply by a constant
factor 2, we are left with
\begin{equation}
\tilde{U}(z,t) = A\, e^{i(\omega t - k_z z)} + B\, e^{i(\omega t +
k_z z)}.
\end{equation}
The real-valued signal $U(z,t)$ has been transformed into the
complex-valued signal $\tilde{U}(z,t)$ such that $U(z,t) =
Re(\tilde{U}(z,t))$. With that complex signal at hands, it is now
easy to separate the two wave beams by looking at the Fourier
transform in space
\begin{equation}
\tilde{U}(z,t) = (A e^{i\omega t})\, e^{-ik_z z} + (B e^{i\omega
t})\, e^{ik_z z}.
\end{equation}
Isolating the positive (resp. negative) values of the
wavenumber $k_z$ will isolate the wave propagating towards positive
(resp. negative) $z$. It is important to stress that this second
stage is only possible because $\tilde{U}$ is a complex valued
signal and not a real valued one: its Fourier transform is therefore
not the sum of two complex conjugated parts on positive and negative
frequencies.

\subsection{The two-steps procedure for two-dimensional waves}
Let us now precise how we operate on real experimental data
involving two spatial dimensions.

The first step, called {\em demodulation}, is obtained by performing
sequentially the three following operations:\begin{itemize}
\item[i)] a Fourier transform in time of the field $U(x,z,t)$,
\item[ii)] a wide or selective band-pass filtering in Fourier space around the positive
fundamental angular frequency $\omega=2\pi f$, where we have
introduced $f$ the temporal frequency measured in hertz.
\item[iii)] the inverse Fourier transform generating the complex signal $\tilde{U}(x,z,t)$.
\end{itemize}
On step ii) which removes exactly half the energy of the signal, we
also perform a multiplication by a factor 2, to preserve the
amplitude of the signal and to have $U = Re(\tilde{U})$.

It is crucial to realize that four different travelling waves are
mixed in this complex signal
\begin{equation}\label{eq_classification}
\tilde{U}(x, z, t) = \tilde{A}(x,z,t) + \tilde{B}(x,z,t) +
\tilde{C}(x,z,t) + \tilde{D}(x,z,t)
\end{equation}
with
\begin{eqnarray}
\tilde{A}(x,z,t) &=& A(x,z,t)\, \exp{i(\omega t - k_xx - k_zz)} \,, \label{equation6}\\
\tilde{B}(x,z,t) &=& B(x,z,t)\, \exp{i(\omega t - k_xx + k_zz)} \,, \label{equation6a}\\
\tilde{C}(x,z,t) &=& C(x,z,t)\, \exp{i(\omega t + k_xx - k_zz)} \,, \label{equation6b}\\
\tilde{D}(x,z,t) &=& D(x,z,t)\, \exp{i(\omega t + k_xx + k_zz)} \,. \label{equation6c}
\end{eqnarray}
Note that in Eqs~(\ref{equation6})-(\ref{equation6c}), we have
considered the wavenumbers $k_x$ and $k_z$ positive in order to
identify more easily the direction of propagation.

\begin{figure}
\begin{center}
\includegraphics[width=0.45\linewidth]{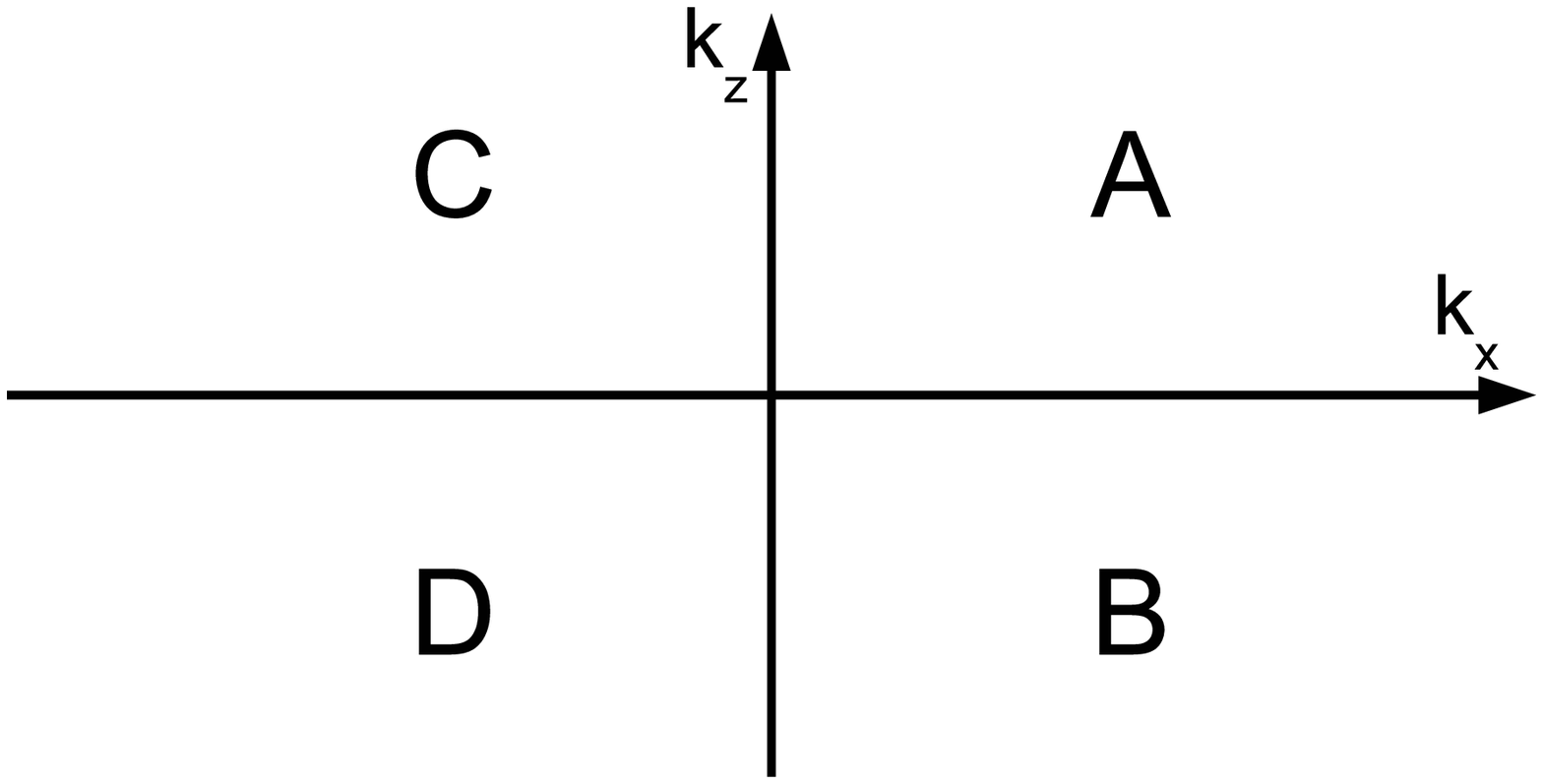}
\includegraphics[width=0.38\linewidth]{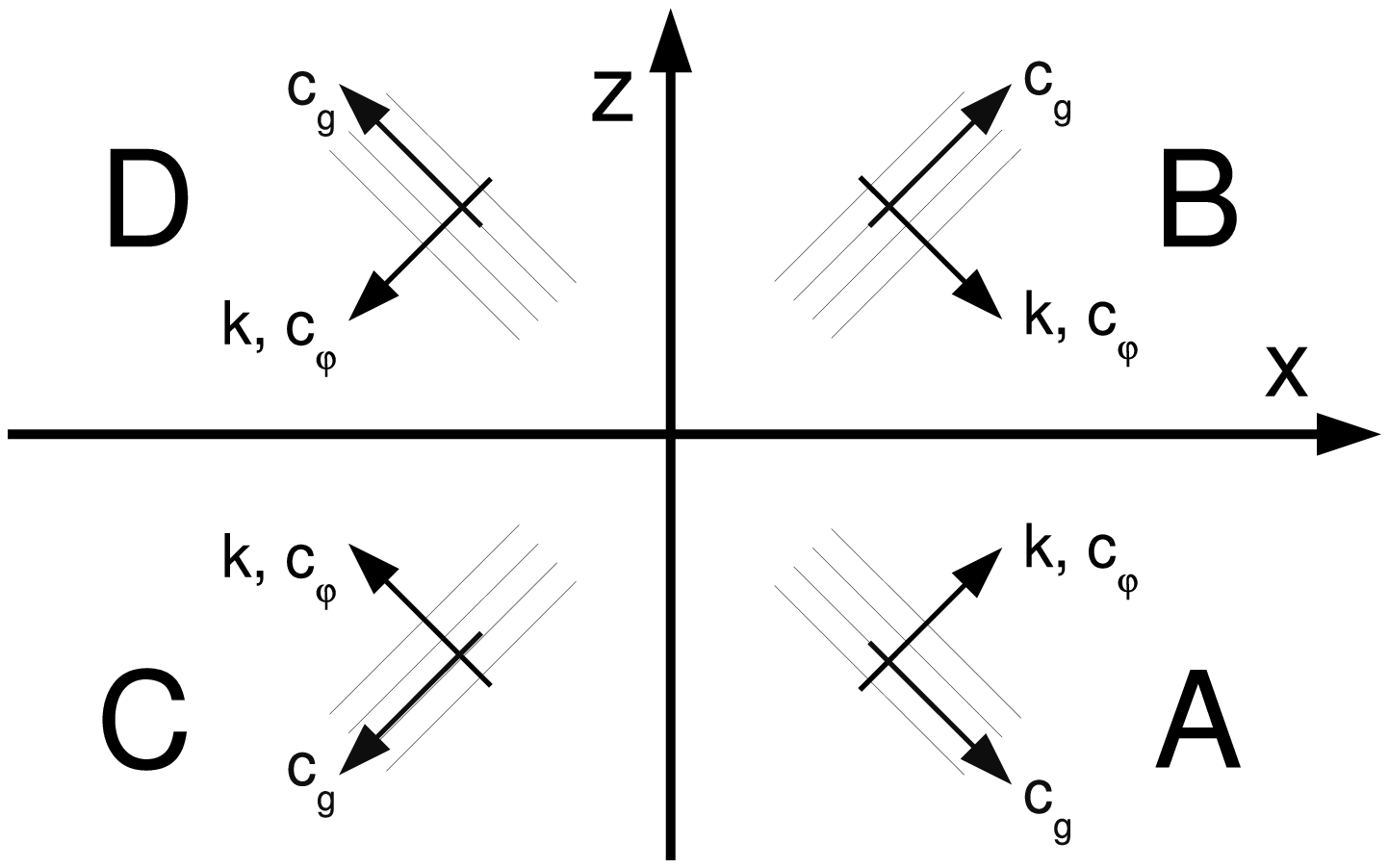}
\end{center}
\caption{These pictures emphasize the four different travelling
waves which might propagate in a stratified fluid. Left panel
corresponds to the wavenumbers in the Fourier space while the right
panel is in the direct space. Left panel defines the four different
domains $A$, $B$, $C$ and $D$ corresponding to different signs of
the wavenumber in the $x$ and $z$ space directions.  Right panel
presents also a synthetic view of the four internal wave beams
emitted, in a 2D stratified fluid, by a vertically oscillating body
located at the origin. For each beam, the phase velocity
$\mathbf{c}_\varphi$ is parallel to the wavevector~$\mathbf{k}$, but
orthogonal to its associated group velocity $\mathbf{c}_g$. Note
that, for example, the beam propagating in the bottom-right domain
of the right panel, corresponds to a wavevector with both positive
components, explaining that this domains is labelled $A$, according
to~(\ref{equation6}).} \label{fig:Fourier}
\end{figure}

Although the four waves oscillate in time at the same frequency
$\omega$, they do not propagate in the same direction, because of
the different signs in front of the wavenumbers $k_x$ and~$k_z$
({\em cf.} Fig.~\ref{fig:Fourier}). Note that amplitudes $A$, $B$,
$C$ and $D$ might depend on space and time: dissipation is a good
example. However, scales on which they vary must be much larger than
scales $\omega^{-1}$, $k_x^{-1}$ and $k_z^{-1}$ around which the
demodulation is performed.

In the second step, we isolate the four waves $A$, $B$, $C$ and $D$,
one from each other, using the complex-valued field $\tilde{U}(x, z,
t)$. To do so, we apply another filtering operation in Fourier
space, but this time in the wavenumber directions $k_x$ and $k_z$
associated with spatial directions $x$ and $z$. Again, this
filtering is only possible on a complex field, {\em i.e.} after the
Hilbert transform has been performed. The goal of this additional
filtering is only to select positive or negative wavenumbers, but
one might also take advantage to apply a more selective filter to
remove spurious details and noise at other wavenumbers.

The two steps we have detailed involved successively a Fourier
transform in time, and then in space directions. Of course, it is
equivalent to operate first in a space direction and then in time
and in the other space direction. The best choice is in fact imposed
by the Fourier transforms resolution, {\em i.e.} the first Fourier
transform has to be performed in the direction with the largest
number of experimental points.

According to the schematic figure~\ref{fig:Fourier}, we then get a
single wave corresponding to a specific direction of the wavevector
where $A(x,z,t)$ is the complex-valued amplitude of the wave
travelling to the right in the $x$-direction and travelling up in
the $z$-direction. $B$, $C$ and $D$ are the complex-valued
amplitudes of the three other possible waves. In the experiments,
one has to measure first the frequency $\omega$ and the wavenumbers
$k_x$, $k_z$, but we note that they are the same over all
spatiotemporal data corresponding to a given experiment. The
envelopes $A$, $B$, $C$ and $D$ contain information not carried by
the fast frequencies $\omega$ and fast wavenumbers $k_x,k_z$, such
as amplitude envelopes of the beams, and local wavenumber
modulations: we will study carefully these quantities.

In summary, the demodulation technique extracts from the
experimental signal the complex quantity
\begin{equation}
\chi(x,z,t) = |\chi(X,Z,T)| \exp [i\varphi_\chi(x,z,t)] \,,
\end{equation}
where $\chi$ stands for $A$, $B$, $C$ or $D$. The argument of the
exponential,  $\varphi_\chi$, is the fast-varying phase
corresponding to wave $\chi$, rotating at the experimental signal
frequency, while containing slow modulations.

In practice, the complex demodulation of the initial spatio-temporal
signal $U(x,z,t)$ results in four sets of four fields (local and
instantaneous):\begin{itemize}
\item[i)] the
amplitude $|\chi(x,z,t)|$,
\item[ii)] the frequency $\omega(x,z,t)=\partial \varphi_\chi / \partial t$,
\item[iii)] the wavenumber in the $x$-direction
$k_x(x,z,t)=\partial \varphi_\chi / \partial x$,
\item[iv)] the wavenumber in the $z$-direction $k_z(x,z,t)=\partial \varphi_\chi /\partial z$.\end{itemize}
Note that the wavenumbers and the frequency have to be calculated
from the phase field.

Let us emphasize finally that, the Fourier transform being bijective
only when applied to infinite or periodical signals, it is important
to filter the data first in time, in order to benefit from the
sharpness of time spectra obtained after long-time data
acquisitions; in a second step, the Fourier transform in space is
applied, allowing to separate waves $A$, $B$, $C$ or $D$.

The application of the Hilbert Transform (HT) to the study of
internal waves can provide very interesting results and answer
questions that remained unsolved. The main idea is to isolate the
differences between internal wave beams propagating up or down, to
the left or to the right. However, before considering such
situations, we study in the following subsection how this method
might be applied to a simple two-dimensional situation which has
been intensively studied already.

\subsection{The classical oscillating cylinder experiment as a first
example} \label{Applications}

The first example one might consider is the simple experiment of a
cylinder oscillating up and down at a given frequency $\omega$.
Initiated by the G\"ortler experiment~\cite{Gortler1943}, this setup
was later popularized by Mowbray and Rarity~\cite{Mowbray1967} and
recently generalized to a three dimensional
situation~\cite{Onu2003}.

The experiment we will describe was realized in a tank of
$120\,\textrm{x}\,50\,\textrm{x}\,10\,\textrm{cm}^3$ filled with
linearly stratified salt water. Quantitative internal waves
visualization were obtained by Synthetic
Schlieren~\cite{Dalziel2000}, which measures the horizontal and
vertical density gradient perturbations referred as $\rho_x(x,z,t)$
and $\rho_z(x,z,t)$ in the following. If one considers an
oscillating cylinder in a two-dimensional stratified fluid, the four
internal wave beams emitted have four wavevectors differing one from
each other by the sign of their projections onto $(Ox)$ and $(Oz)$,
as summarized in Fig.~\ref{fig:Fourier}(b).

\begin{figure}[htb]
\begin{center}
\includegraphics[width=\linewidth]{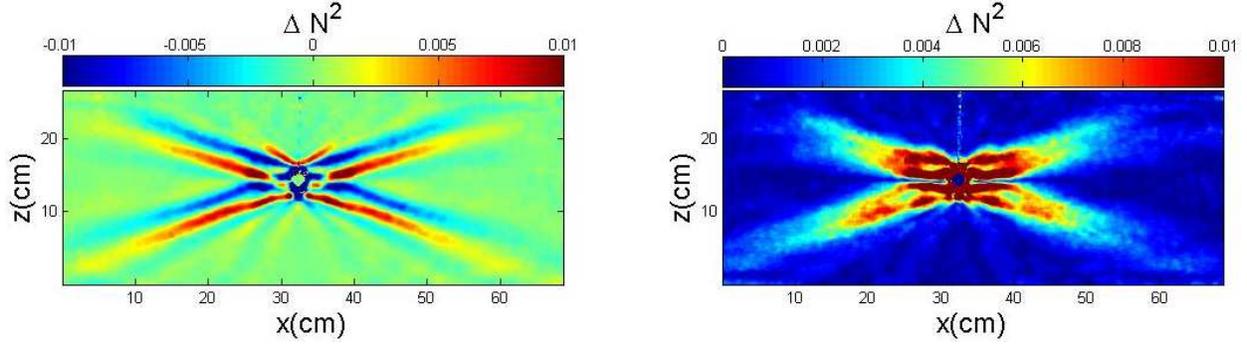}
\caption{Saint Andrew's cross obtained with a cylinder of radius
$R=1.5\,$cm oscillating vertically with an amplitude $1.5\,$mm at a
frequency $\omega=0.28\,\textrm{rad.s}^{-1}$ in a stratified fluid
with Brunt-V\"ais\"al\"a frequency $N=1.0\,\textrm{rad.s}^{-1}$. The
picture presents the real part (left panel) and amplitude (right
panel) of the Hilbert transform, corresponding to the experimental
horizontal density gradient $\rho_x$ filtered in time.}
\label{fig:Hilbert:step1}
\end{center}
\end{figure}

The complex demodulation of the wave field, in time first, is
presented in Fig.~\ref{fig:Hilbert:step1}. Such a picture clearly
emphasizes that four different beams are generated by the
oscillating cylinder, all of them being tilted with an angle
$\theta$ with respect to the gravity, $\theta$ being given by the dispersion
relation~(\ref{dispersionrelation}).

\begin{figure}[htb]
\begin{center}
\includegraphics[width=\linewidth]{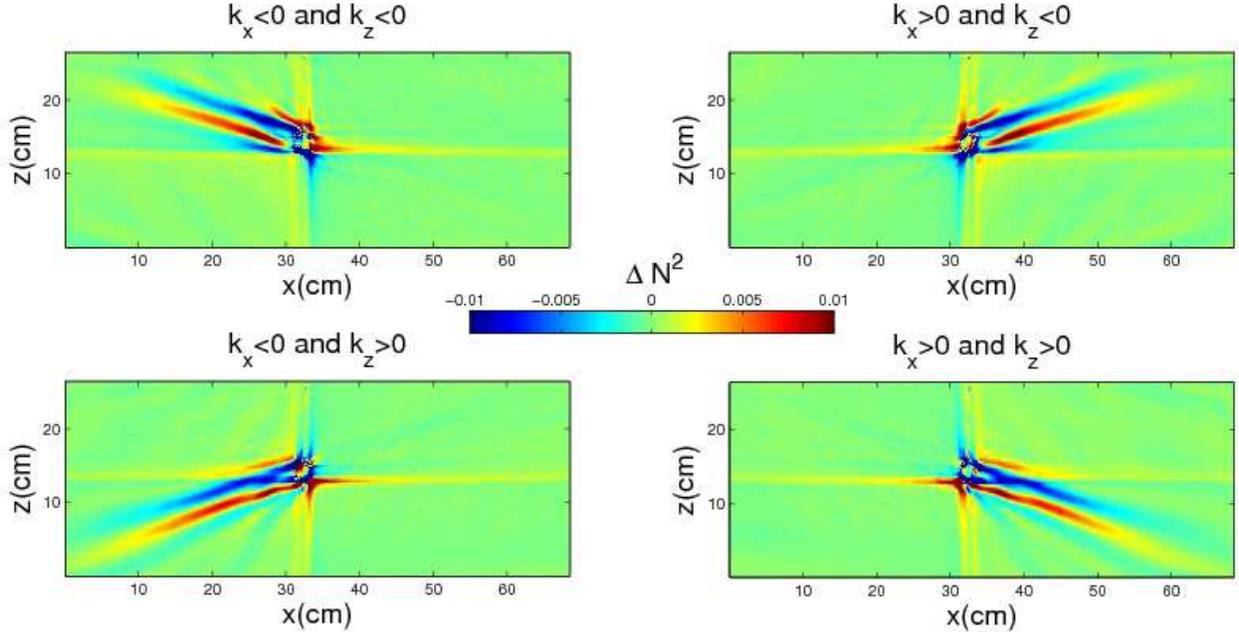}
\caption{Real part of the horizontal density gradient $\rho_x$
presented in Fig.~\ref{fig:Hilbert:step1} after spatial filtering.
The four different pictures correspond to the four possible waves
described in Fig.~\ref{fig:Fourier}(a).} \label{fig:Hilbert:step2}
\end{center}
\end{figure}

After an additional filtering along  the \mbox{$z$-coordinate},
first, and then along the $x$-coor\-di\-nate, four different beams
can be isolated as presented in Fig.~\ref{fig:Hilbert:step2}.
Although some boundary effects can be detected at locations
corresponding to discontinuities in space due to the cylinder,
explaining the intense yellow areas along the horizontal and
vertical axes, it is important to stress that there is no ambiguity
concerning the field treated. Moreover, these side-effects might be
corrected by applying the HT in space only to a selected domain
instead of considering the full window of observation containing
also the cylinder here.

We will now present two interesting points that have not been
addressed in the previous literature (for recent results
see~\cite{Gostiaux_these,Zhang2007}) while studying the wavefield
emitted by an oscillating cylinder.

\begin{figure}
\begin{center}
\includegraphics[width=0.6\linewidth]{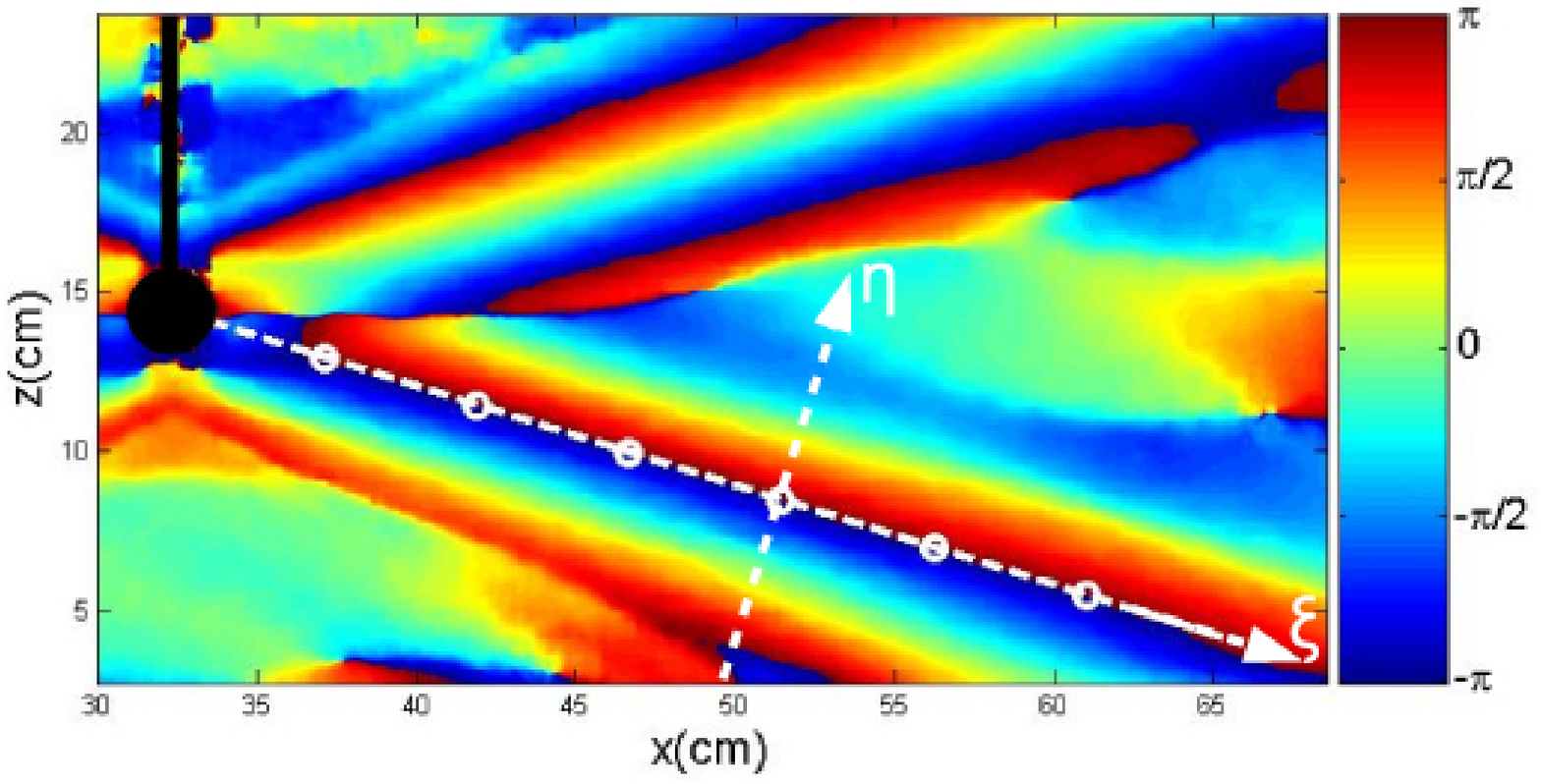}
\includegraphics[width=0.38\linewidth]{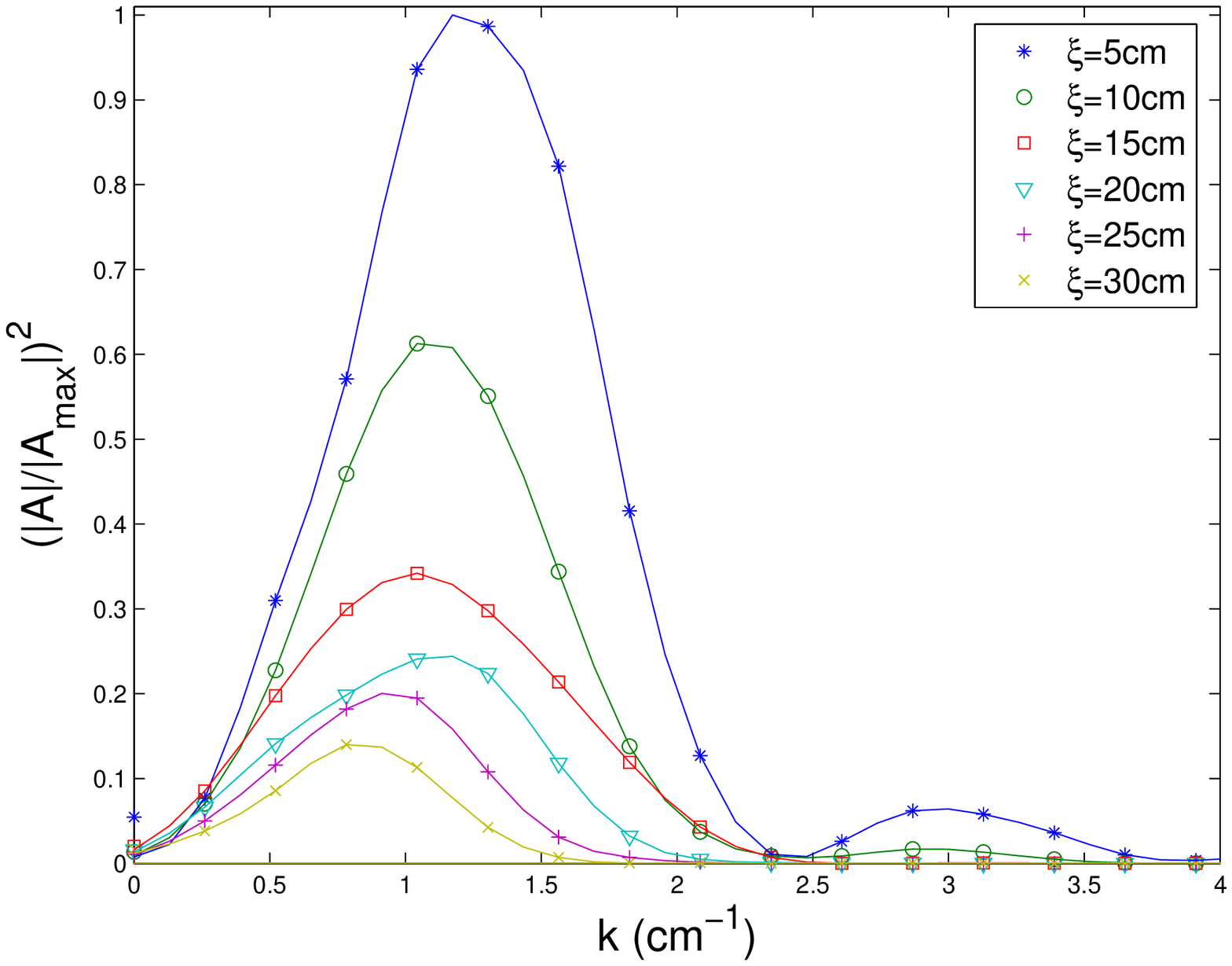}
\end{center}
\caption{(\textit{Left panel}) Phase of the density gradient
$\rho_x(x,z,t)$ after step 1, zoomed on the right-hand side of
Fig.~\ref{fig:Hilbert:step1}. The cylinder is represented in black.
Note also the definition of the variables $\xi$ and $\eta$,
respectively along and transversal to the propagation.
(\textit{Right panel}) Evolution of the transverse spatial spectrum
along the axis of propagation $(O\xi)$. Amplitudes have been
normalized by the maximum value of the spectrum closest to the
cylinder at $\xi=5$~cm. } \label{fig:PhaseCylinder}
\end{figure}

Figure~\ref{fig:PhaseCylinder}(a) presents a zoom on the phase of
the beams emitted to the right of the cylinder (right side of
Fig.~\ref{fig:Hilbert:step1}): it is clear that there is no direct
link between the phase evolution of the downward and upward
propagating waves. Such an image will be very helpful when we will
analyze the spatial structure of the emitted phase for the
diffraction phenomenon in section~\ref{diffraction}.

Figure~\ref{fig:PhaseCylinder}(b) shows the evolution of the
transverse spatial spectrum of the downward propagating wave to the
right. They have been obtained by extracting the transverse profiles
(along $(O\eta)$) at the circles located on the axis of propagation
of the wave $(O\xi)$ and shown in Fig.~\ref{fig:PhaseCylinder}(a).
This picture reveals not only the decrease of the amplitude due to
dissipation (see Sec.~\ref{dissipation} for a complete analysis) but
also the gradual shift toward smaller value of the wavenumbers, {\em
i.e.} toward larger wavelengths.~\cite{Hazewinkel2007}

In summary, the use of the Hilbert transform allows one to separate
rather easily all the waves emitted from the cylinder, and to have a
very precise definition of the phase of the wavefield, a quantity of
importance to describe the wave spectra. We use in the next section
these properties to address questions still pending.

\section{Applications}
\label{newapplications}

In the remainder of the article, we study internal waves beam
emanating from a ``pocket size"
version~\cite{MercierGostiauxDauxois} of the internal plane wave
generator that we have recently developed~\cite{Gostiaux2007}. All
experiments were realized in a tank of
$80\times42.5\times17\,\textrm{cm}^3$ filled with linearly
stratified salt water. Horizontally oscillating plates of thickness
$6\,\textrm{mm}$ create a sinuso\"idal envelope of amplitude
$a_{0}=5\,\textrm{mm}$ and
wavelength $\lambda_{e}=3.9\,\textrm{cm}$, ($k_{e}=2\pi/\lambda_{e}$). 
The oscillating frequency $\omega$ defines through the dispersion
relation~(\ref{dispersionrelation}) the angle of propagation of the
beam with respect to the gravity. Such a device was shown to be
extremely effective to generate nice plane wave beams in linearly
stratified fluids~\cite{Gostiaux2007}.

\subsection{Dissipation of internal waves}\label{dissipation}

The first physical situation we consider is the dissipation of
internal waves within a laboratory tank. The linear viscous theory
developed by Thomas~\&~Stevenson \cite{Thomas1972} first, and
Hurley~\&~Keady \cite{Hurley1997part2} afterwards, has been tested
with good accuracy \cite{Sutherland1999, Sutherland2002, Zhang2007}.
The damping of the averaged spectrum with time has also been
studied, typically in the case of attractors because a steady state
is obtained due to a balance between amplification at reflection and
viscous damping \cite{Rieutord2001, Hazewinkel2007}. In
Fig.~\ref{fig:PhaseCylinder}(b), the damping of the spectrum along
the axis of propagation can also be analyzed similarly.
Nevertheless, these approaches are integral ones over all
wavenumbers, and the viscous damping has not been tested on a
monochromatic internal wave. The Hilbert Transform is moreover an
excellent tool to measure the dissipation effects.

The structure expected \cite{Lighthill} for a viscous
internal plane wave is
\begin{equation} \psi(\xi,\eta,t) =
\psi_{0}\,e^{-\beta\,\xi}\,e^{i(\omega t - k\eta)}\, ,
\end{equation} where $\xi$ is the
longitudinal coordinate while $\eta$ corresponds to the transversal
one. The quantity
\begin{equation}\beta = \frac{\nu k^3}{2N\cos\theta} =
\frac{\nu k^3}{2N\sqrt{1-\frac{\omega^2}{N^2}}}\, .
\end{equation}
corresponds to the inverse dissipation length. Thanks to the
analytical representation of the internal waves using the HT, it is
easy to get the envelope of a monochromatic internal wave and thus
quantify how it decreases through viscous dissipation. Results shown
below correspond to three different stratifications.

\begin{figure}[htb]
\begin{center}
\includegraphics[width=0.48\linewidth]{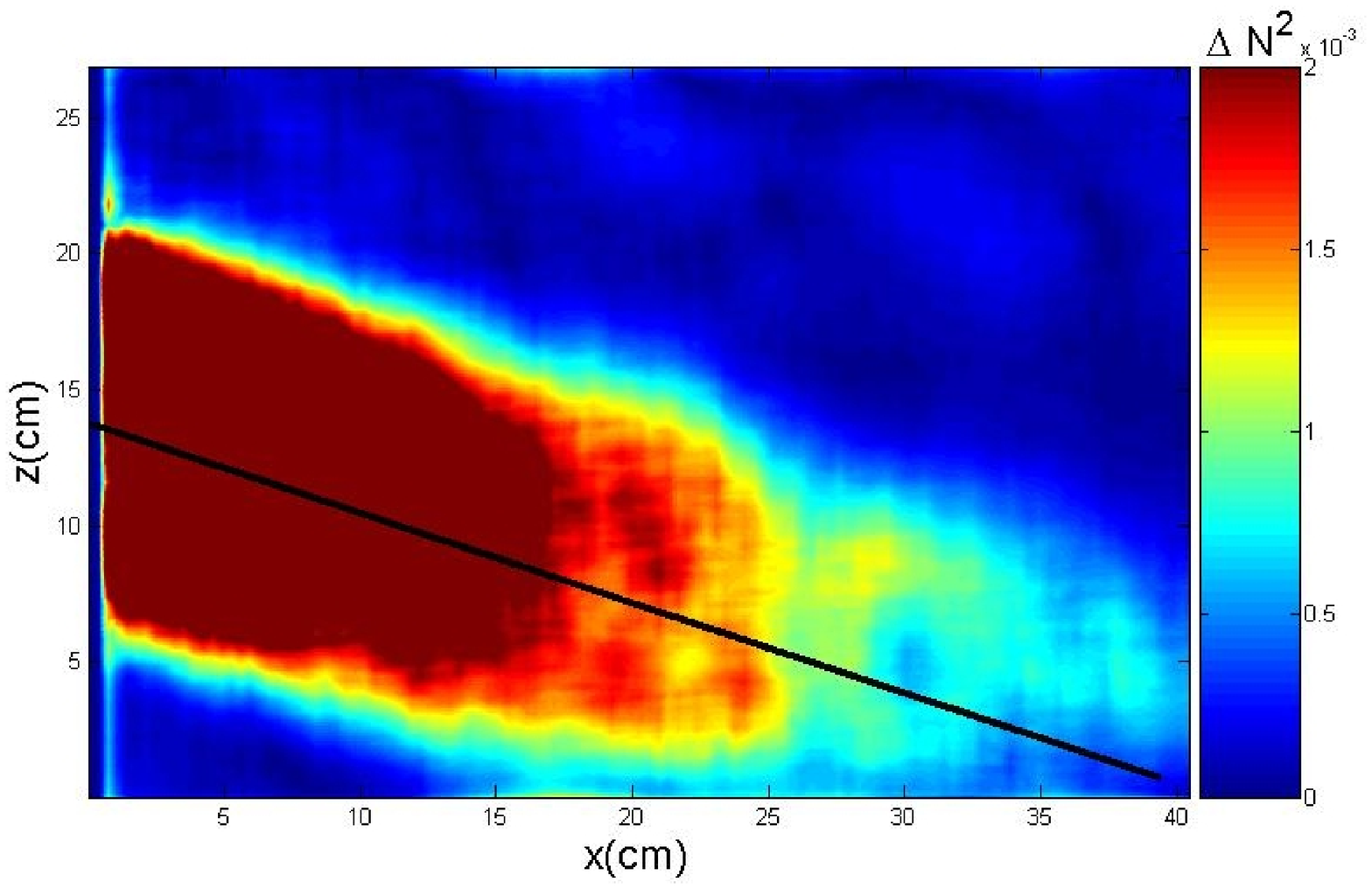}
\includegraphics[width=0.47\linewidth]{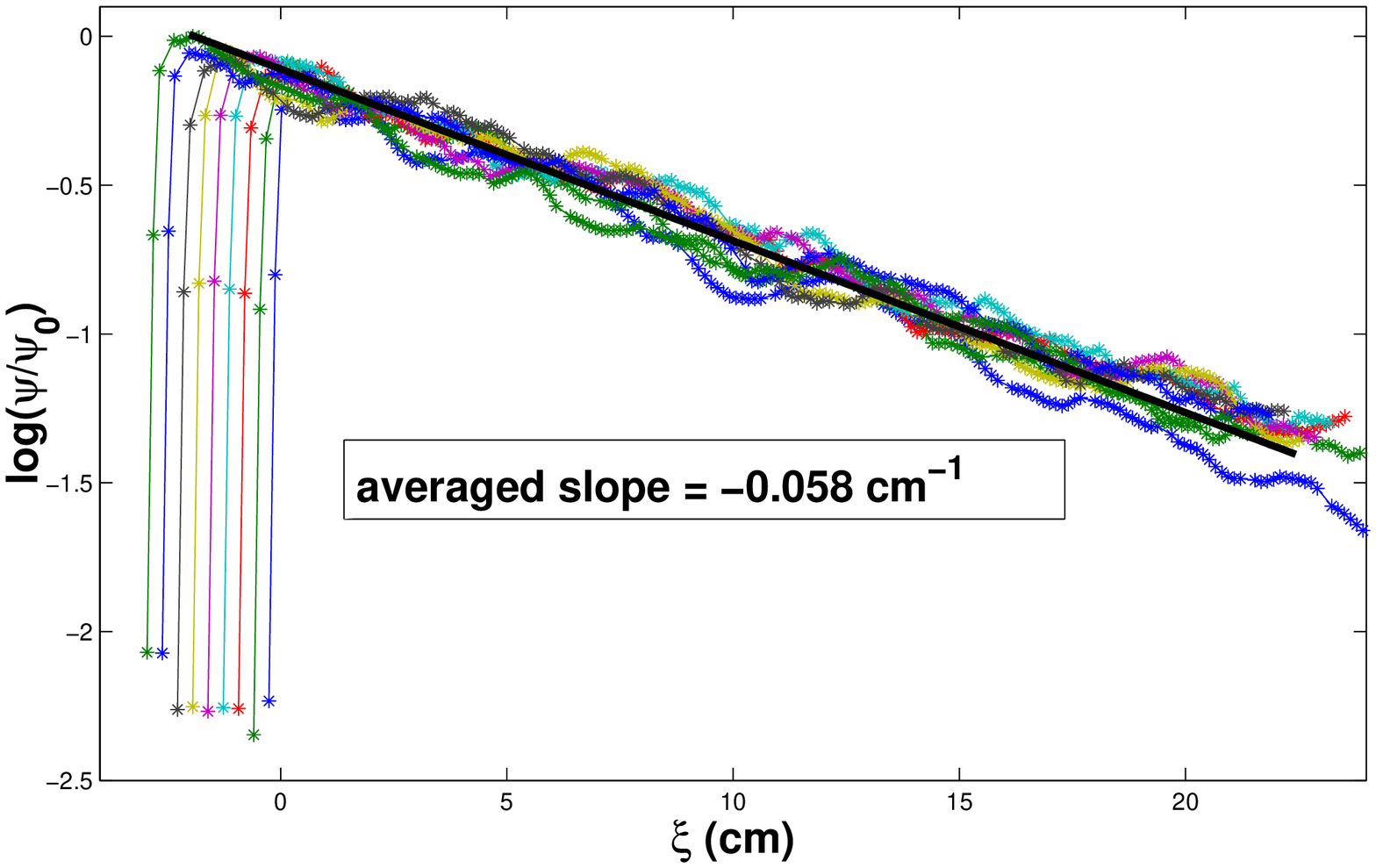}
\caption{(a) Envelope of the Hilbert transform of the downward field
$\rho_x(x,z,t)$ ($A$ and $C$) for an oscillating frequency
$f=0.033\,$Hz and a Brunt-V\"ais\"al\"a frequency
$N=0.66\,\textrm{rad.s}^{-1}$. The tilted black line indicates the
location of one of the regularly extracted profiles. All other ones
are parallel to this one. Panel (b) presents the logarithm of
extracted profiles together with the averaged linear fit.}
\label{envelopesui}
\end{center}
\end{figure}

For each frequency, the envelope of the emitted beam is extracted: a
typical result is shown in Fig.~\ref{envelopesui}(a). The logarithm
along the \mbox{$\xi$-coordinate} is then plotted versus the
longitudinal coordinate $\xi$ for different $\eta$-values, as
illustrated in Fig.~\ref{envelopesui}(b). The dissipation rate
according to the direction of propagation is then obtained by the
averaged linear fit over the different profiles extracted.

\begin{figure}[htb]
\begin{center}
\includegraphics[width=0.5\linewidth]{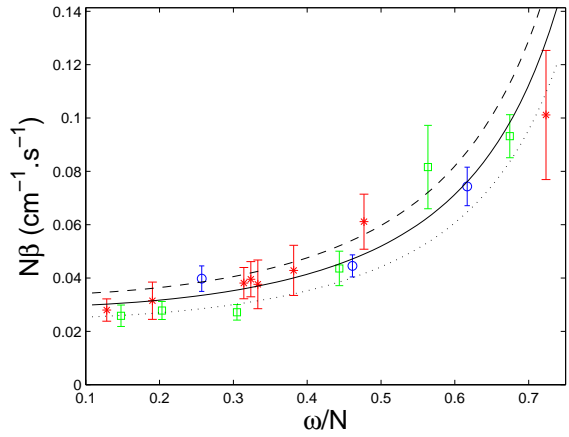}
\caption{$N\beta$ as a function of $\omega/N$ for three different
stratifications: stars, squares and circles correspond to experiment
with $N=0.66$, 0.68 and 0.76 rad.s$^{-1}$. The solid curve
corresponds to the best value for fitting the data, $\lambda_{e}=
3.55$ cm, while the dashed (resp. dotted) line to the lower (resp.
upper) bound $\lambda_e=3.39$ cm (resp. $\lambda_e=3.75$ cm).}
\label{attenuationrate}
\end{center}
\end{figure}

Repeating the above procedure for several frequencies, one gets the
evolution of the dissipation length $\beta(\omega)$. It is however
important to realize that, the propagation being tilted with respect
to the vertical plane of emission, the forcing of the internal plane
wave generator does not create a wave whose wavelength is
$\lambda_e$. A projection of the wavelength on the direction
perpendicular of propagation has to be taken into account: the
wavevector of the propagating internal wave is therefore
$k=k_{e}/\cos\theta$, so that we have the following relation
\begin{equation}
\beta = \frac{\nu k_{e}^3}{2N\cos^4\theta}, \end{equation} which can
be rewritten in the more convenient form
\begin{equation} N\beta = \frac{\nu k_{e}^3}{2}\frac{1}{(1-x^2)^2}
\end{equation}
by introducing $x=\omega/N$. It is thus generic to plot the
attenuation rate $\beta$ times the Brunt-V\"ais\"al\"a frequency $N$
as a function of the ratio $x=\omega/N$ for different values of $N$,
as presented in Fig.~\ref{attenuationrate}. Using the value of the
viscosity $\eta=1.05\,10^{-6}$ m$^2$.s$^{-1}$, only one free
parameter is remaining, the wavevector $k_e=2\pi/\lambda_e$.

Above procedure leads to the following result, \mbox{$\lambda_{e}=
3.55$~cm (+0.20/-0.16)} cm, in good agreement  with the value
obtained from the transverse beam
structure~\cite{MercierGostiauxDauxois}. Surprisingly these values
are slightly different from the one imposed by the internal plane
wave generator. Figure~\ref{attenuationrate} which presents the best
fit attests the good agreement with experimental results. The model
seems particularly accurate for frequencies $\omega$ sufficiently
small compared to the cut-off frequency~$N$.

\subsection{Back-reflected waves on a slope.}
\label{backreflection}

\begin{figure}[htb]
\begin{center}
\includegraphics[width=0.45\linewidth]{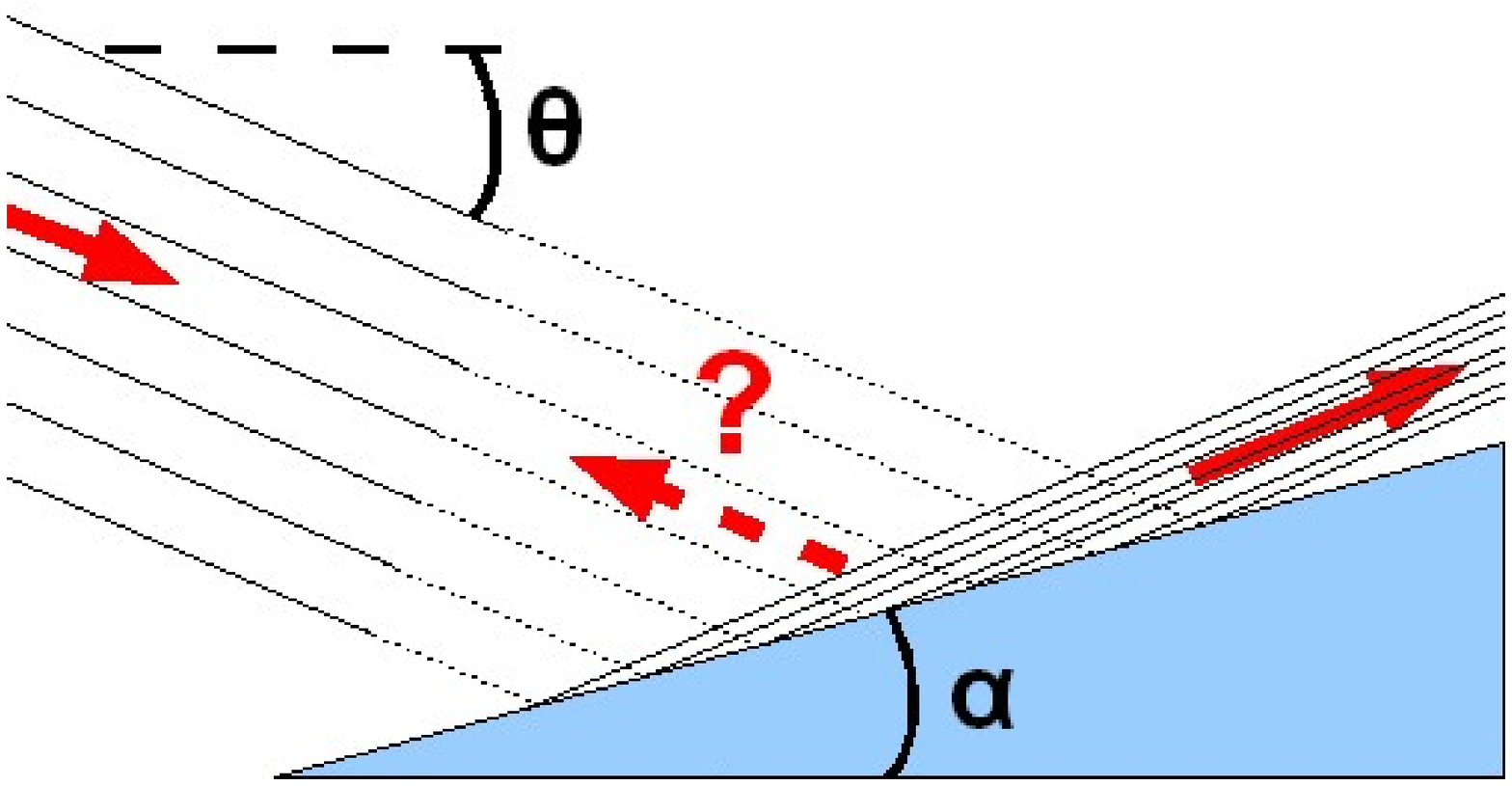}
\includegraphics[width=0.45\linewidth]{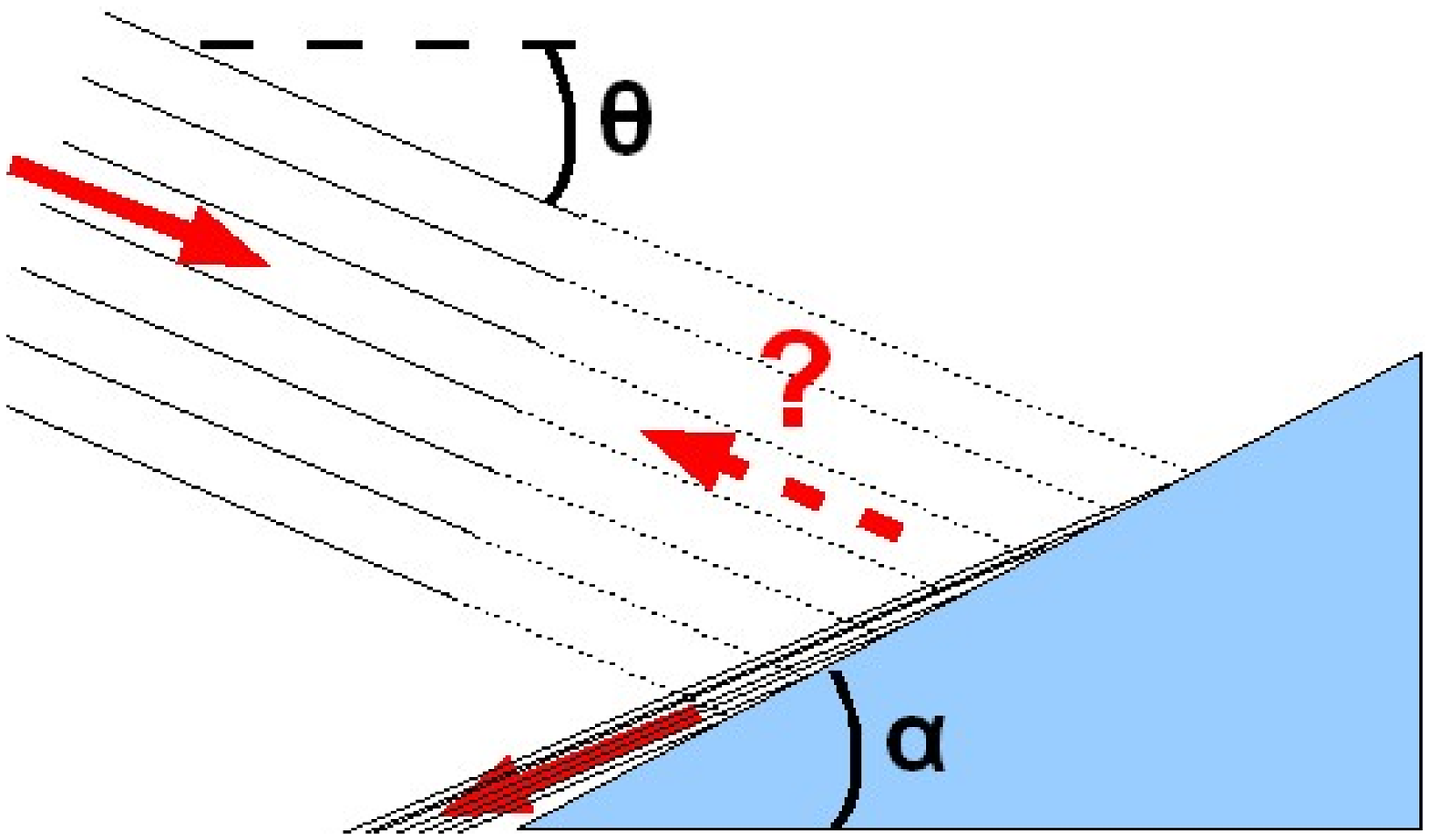}
\caption{Left panel presents the principle of the possible
back-reflection problem for an incident internal wave beam. Left
panel shows the $\theta>\alpha$ case, while the right one presents
the opposite case $\theta<\alpha$.}\label{figureprincipereflection}
\end{center}
\end{figure}

We have also used the Hilbert Transform to identify a possible
back-reflected wave when an incident internal wave beam is
reflecting on a slope of angle $\alpha$ with the horizontal. After
reflection, as shown by Fig.~\ref{figureprincipereflection}, two
beams inclined with an angle $\theta$ with respect to the horizontal
might be emitted from the slope. One of this beam has been
experimentally reported several
times~\cite{Sutherland1999,PeacockTabei,Gostiaux_these,NyeDalziel},
contrary to the second one, aligned with the incident beam but
propagating in the opposite direction and represented by the dashed
arrow in Fig.~\ref{figureprincipereflection}. This additional beam
was considered by Baines~\cite{Baines1971part1} and
Sandstrom~\cite{Sandstrom} when studying theoretically the effect of
boundary curvature on reflection of internal waves. Let us
experimentally prove that no back-reflection occurs at planar
surfaces.

\begin{figure}[htb]
\begin{center}
\includegraphics[width=\linewidth]{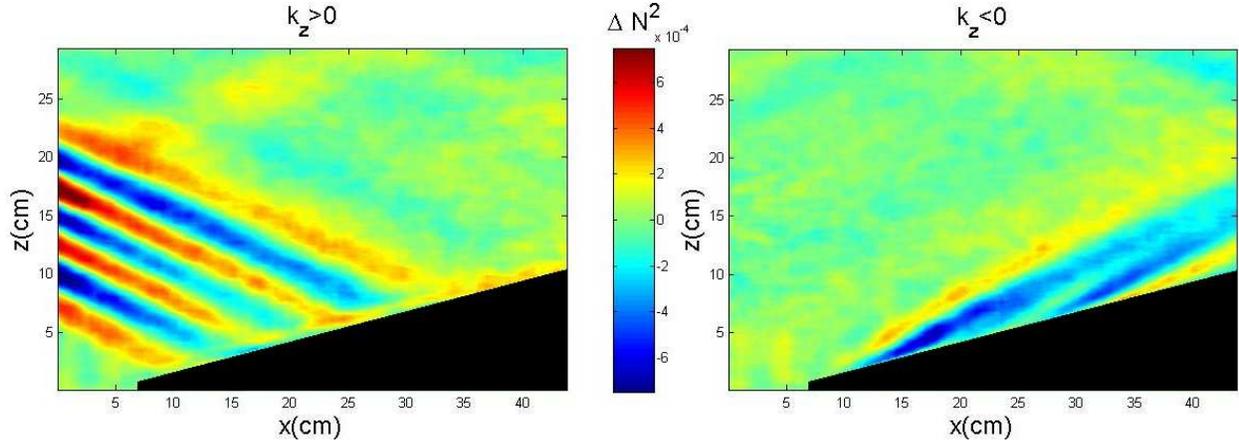}
\caption{Horizontal density gradient $\rho_x$. (Left) Downward ($A$
and $C$) and (right) upward ($B$ and $D$) waves reflecting on a
slope ($\alpha=14°$), in $\Delta N^2$ rad$^2$.s$^{-2}$. The
frequency of the waves was $\omega/N=0.43$ with the
Brunt-V\"ais\"al\"a frequency $N=0.42~\textrm{rad.s}^{-1}$.
}\label{fig_backreflected1}
\end{center}
\end{figure}

It is clear that if it exists the amplitude of the back-reflected
beam has to be much smaller than the incident one, as usual
techniques were unable either to identify it, or to exclude it. This
is the reason why we have performed several experiments of an
incident beam impinging onto a slope, away from critical incidence
but also close to it (see Table~\ref{table:conditions} for values of
control parameters).
Analysis of one case with $\theta>\alpha$ is presented in
Fig.~\ref{fig_backreflected1}. The back reflected beam in that case
should be a $D$-wave according to
classification~(\ref{eq_classification}). However,
Fig.~\ref{fig_backreflected1} shows absolutely no evidence of it,
and only a $B$-component is visible. Nevertheless, as the HT along
the $x$-coordinate has not been performed to avoid the introduction
of spurious boundary effects, it is still possible to argue that the
$D$-wave might be localized where the $B$-wave could shadow it.
However, as this spatial domain remains extremely small, it
is therefore very unlikely.

\begin{table}[!ht]
\centering
\begin{tabular}{|l|r @{.} l|r @{.} l|r @{.} l|r @{.} l|r @{.} l|}
\hline
Run & \multicolumn{2}{c}{1}\vline & \multicolumn{2}{c}{2}\vline & \multicolumn{2}{c}{3}\vline & \multicolumn{2}{c}{4}\vline & \multicolumn{2}{c}{5}\vline \\
\hline
$\theta$ (deg.) & 14&0 & 7&0 & 11&4 & 15&1 & 25&0 \\
\hline
$\alpha$ (deg.) & 25&5 & 14&5 & 14&5 & 14&5 & 14&0 \\
\hline
$\varepsilon$ (deg.) & -11&5 & -7&5 & -3&1 & 0&6 & 11&0 \\
\hline
$N$ (rad.s$^{-1})$ & 0&42 & 0&58 & 0&58 & 0&58 & 0&42 \\
\hline\end{tabular} \caption{Summary of experimental runs with all
control parameters:  $\theta$ the angle of energy propagation,
$\alpha$ the angle of the slope, $\varepsilon=\theta-\alpha$ and $N$
the Brunt-V\"ais\"al\"a frequency.}\label{table:conditions}
\end{table}

Varying the angle $\theta$ of the waves around $\alpha$, the slope
angle, no trace of back-reflected intensity are apparent, even close
to critical conditions $\theta=\alpha$. In order to give a
definitive answer, we have also considered the case $\theta<\alpha$
(see Fig.~\ref{figureprincipereflection}(b)). In that case, the
back-reflected beam would be the only one to propagate upward, while
the classically reflected beam would be a $C$-wave, propagating
downward.
\begin{figure}[htb]
\begin{center}
\includegraphics[width=\linewidth]{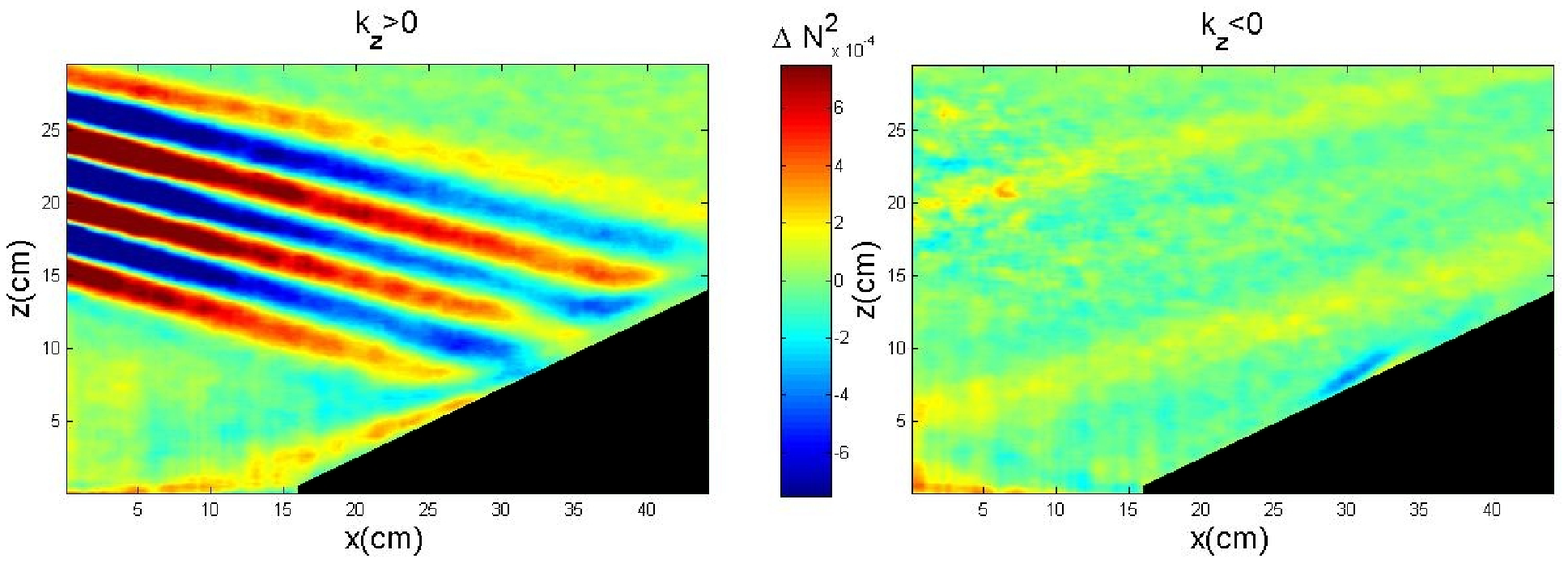}
\caption{Horizontal density gradient $\rho_x$. (Left) Downward waves
($A$ and $C$) and (right) upward waves ($B$ and $D$) reflecting on a
slope ($\alpha=25.5°$), in $\Delta N^2$ rad$^2$.s$^{-2}$. The
frequency of the waves was $\omega/N=0.24$ with the
Brunt-V\"ais\"al\"a frequency $N=0.42~\textrm{rad.s}^{-1}$.
}\label{fig_backreflected2}
\end{center}
\end{figure}
Figure~\ref{fig_backreflected2} corresponding to such a case attests
that there is no wave propagating backward. We can therefore claim
that the back-reflected beam is definitely not present when internal
waves reflect onto a slope. Concave or convex slopes might lead to
different results~\cite{Baines1971part1}.

\subsection{Diffraction of internal waves}
\label{diffraction}

The diffraction of internal waves is the last issue we will consider
in this paper. Although it is not directly interesting for
oceanographic applications, it seems natural to ask~\cite{LeoMaas}
what is the equivalent of the Huygens-Fresnel principle for optical
waves. Indeed, it has been established for centuries that when a
plane wave encounters a thin slit, optical waves are reemitted in
all directions. But what about internal waves? To the best of our
knowledge, there is neither theoretical nor experimental results on
this topic.

\begin{figure}[htb]
\begin{center}
\includegraphics[width=0.4\linewidth]{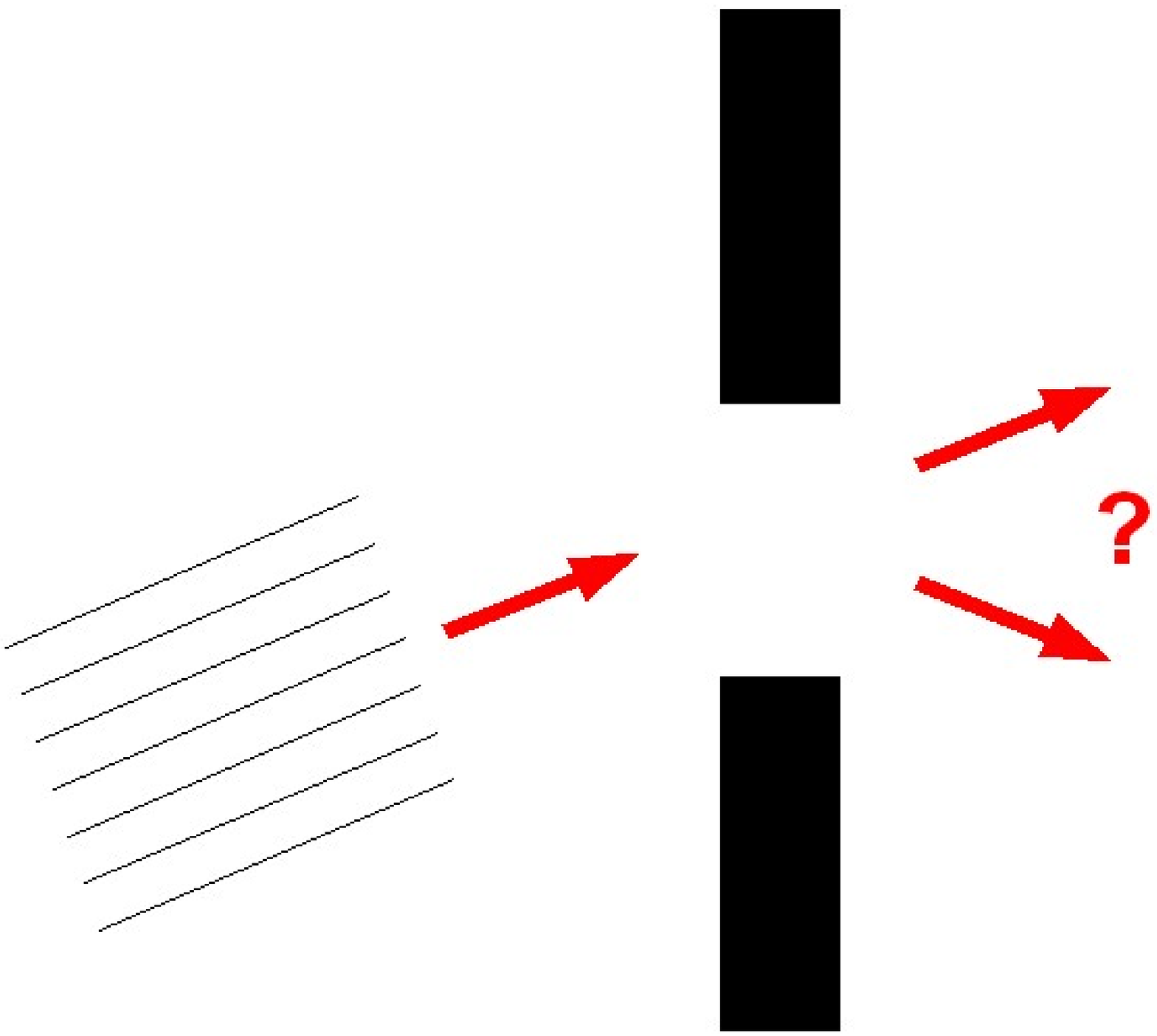}
\includegraphics[width=0.57\linewidth]{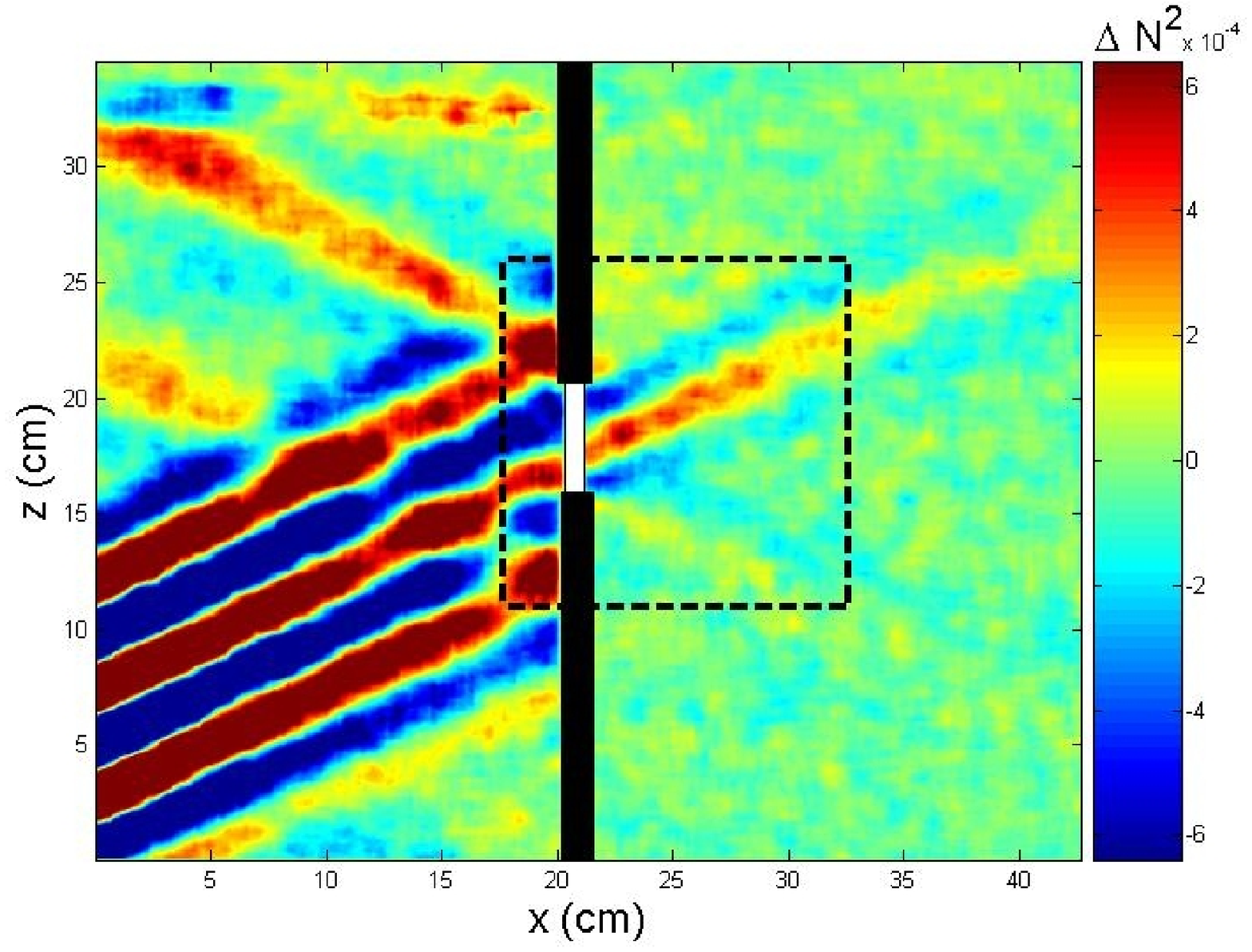}
\caption{Left panel presents the principle of the diffraction
problem for an incident internal wave beam. Right panel shows the
horizontal density gradient $U$ field in $\Delta N^2$
rad$^2$.s$^{-2}$ for an incoming internal plane wave, corresponding
to a wavelength $\lambda=3.6$~cm on a 4~cm wide slit. The dotted
square corresponds to the region presented in Figs.~\ref{fig:a_4cm}
and~\ref{fig:a_2cm}.}\label{figurediffraction}
\end{center}
\end{figure}

As the incident wave is impinging onto the slit with a well defined
frequency, it is clear that the transmitted waves have to satisfy
the dispersion relation~(\ref{dispersionrelation}). However, as
schematically shown in Fig.~\ref{figurediffraction}(a), two
different beams might be expected after the slit. In the case
exemplified in this picture, it is clear that most of the energy
will be transmitted to the waves propagating {\em upward}. Is it
possible however to detect whether part of the energy is emitted
{\em downward}? The main goal is therefore to be able to
discriminate what is going out of a slit with a width comparable to
the wavelength of an incoming internal plane wave.

\begin{figure}[htb]
\begin{center}
\includegraphics[width=\linewidth]{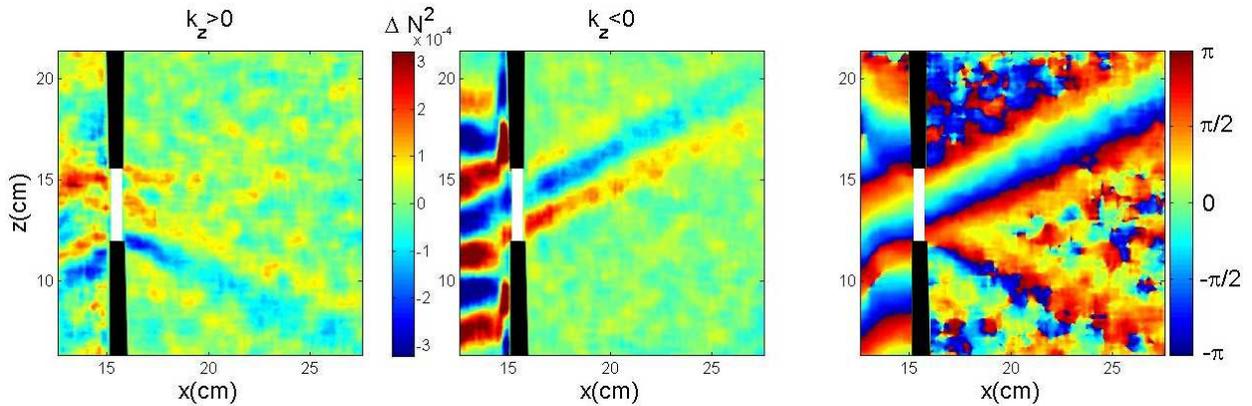}
\caption{{\em Large} slit case. Hilbert transform of the gradient
density field $\rho_x$ in $\Delta N^2$ $(\textrm{rad.s}^{-1})^2$
filtered at $\omega=0.196~\textrm{rad.s}^{-1}$ for $a=4$~cm with
$k_z>0$ (left panel), $k_z<0$ (centered panel) and phase of the
complete field with all values of $k_z$ (right panel).}
\label{fig:a_4cm}
\end{center}
\end{figure}

\begin{figure}[htb]
\begin{center}
\includegraphics[width=\linewidth]{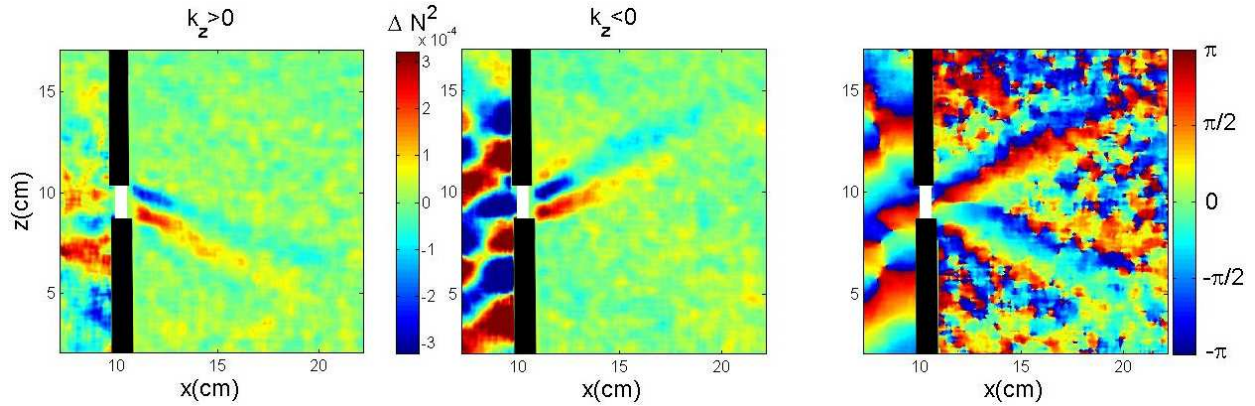}\\
\caption{{\em Small} slit case. Hilbert transform of the horizontal
density gradient field $\rho_x$ in $\Delta N^2$
$(\textrm{rad.s}^{-1})^2$ filtered at
$\omega=0.196~\textrm{rad.s}^{-1}$ for $a=2$~cm with $k_z>0$ (left
panel), $k_z<0$ (centered panel) and phase of the complete field
with all values of $k_z$ (right panel).} \label{fig:a_2cm}
\end{center}
\end{figure}

In th experiments, the stratification is linear with
\mbox{$N=0.45~\textrm{rad.s}^{-1}$}, while the incoming beam has a
frequency $\omega=0.196~\textrm{rad.s}^{-1}$ and a wavelength
$\lambda=3.2$~cm. Since the source is the internal plane wave
generator, we remind that it corresponds to a vertical wavelength
$\lambda_z=3.6$~cm. The slit of varying width $a$ is made of two
sliding plastic plates of a thickness of $1$~cm, and is represented
by a thick vertical white line in Figs.~\ref{figurediffraction},
\ref{fig:a_4cm} and~\ref{fig:a_2cm}, since no signal can be obtained
in this region with the Synthetic Schlieren technique because of the
sides of the slits.  We present below results corresponding to
widths of the slit $a=6$, 4, 3, and 2~cm. Note that for $a=1$~cm, no
signal was obtained after the slit, which means that its intensity
was below the noise level (if there was anything to measure).

Figures~\ref{fig:a_4cm}  and~\ref{fig:a_2cm} present the results for
$a=4$~cm and $a=2$~cm, emphasizing two different mechanisms for the
emission that corresponds to $a>\lambda$ and $a<\lambda$. Note that
both pictures present zooms close to the slit to better appreciate
the interesting region.

In the first case, it is apparent that most of the intensity is in
the beam emitted in the same direction than the incoming wave
(upward propagation here). Furthermore, this beam seems very similar
to the incoming plane wave: one notes indeed that the wavelength is
identical before and after the slit. Moreover the continuity of the
phase is nicely shown by the right panel of Fig.~\ref{fig:a_4cm}.
Nevertheless, the edges of the slit are also a source emitting
downward propagating beams since waves can be seen on both sides of
the slit. It seems logical since the incoming plane wave creates an
oscillating flow close to the slit, inducing a wavefield similar to
the one of an oscillating body in a fluid at rest. It is finally
important to notice that the spatial structure of the phase of the
complete wave field is different from the one observed in
Fig.~\ref{fig:PhaseCylinder} for an oscillating cylinder. The
emission of the upward and downward propagating waves by the slit is
consistent since there is no discontinuity in the spatial structure
of the phase.

In the second case $a=2$~cm presented in Fig.~\ref{fig:a_2cm} with a
slit smaller than the wavelength, the mechanism is different. It
seems that the only property similar to the incoming plane wave in
the two beams transmitted through the slit is the frequency. Both
transmitted beams have comparable intensities. The spatial structure
of the phase presents a discontinuity strongly reminiscent of the
wave field emitted by an oscillating body as shown by
Fig.~\ref{fig:PhaseCylinder}.

\begin{figure}[htb]
\begin{center}
\includegraphics[width=0.5\linewidth]{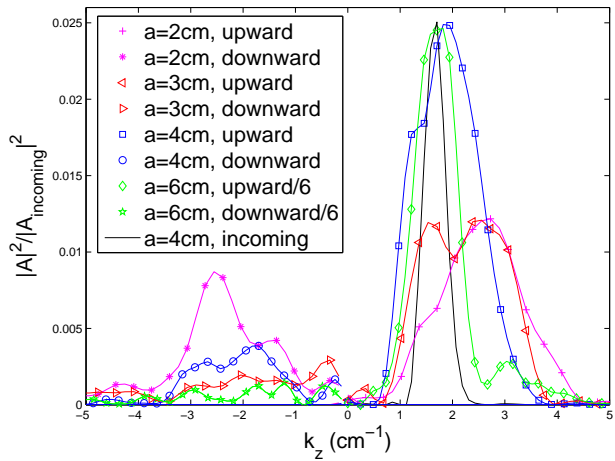}
\caption{Spectra of the vertical cut of the  horizontal density
gradient $\rho_x$ measured $1.5$~cm on the right of the slit. The
different curves corresponds to different widths~$a$ of the slit
(see inset for values).} \label{fig:spectra_after_slit}
\end{center}
\end{figure}

To have a global view of the physics of internal plane waves
diffraction, we finally present in Fig.~\ref{fig:spectra_after_slit}
the vertical spatial spectra associated to the transmitted waves
(upward and downward) taken at 1.5 cm after the slit for all values
of the width $a$, in comparison with the spectrum of the incoming
wave. The amplitudes of the Fourier components have been normalized
by the maximum amplitude $|A_{incoming}|$ of the incoming wave
spectrum measured $4$~cm before the slit. Several comments are in
order. A clear shift of the peak toward larger values of the
wavenumbers is visible when the width of the slit decreases. The
spectra are also clearly enlarged. This is consistent with the
previous remark that for a large slit the transmitted wave beam is
very similar to the incident one. The spectra of the downward beam
are visible in the negative~$k$ half plane. In the large cases,
$a=3$ and 4, they are wide and with a small amplitude attesting that
most of the incident energy is transmitted upward, i.e. directly. On
the contrary, in the thin slit case, $a=2$, the amplitudes for
downward and upward propagating are comparable. It is difficult to
propose a more quantitative discussion since the dissipation of the
spectra is important.

In summary, the analysis of these spectra confirms that when the
slit is sufficiently ``large", the emitted beam has a vertical
wavenumber similar to the incoming one although the spectrum is
slightly wider. On the contrary, when the slit is ``small" enough,
both beams have similar spectra and amplitudes.

Finally, we can conclude that the change in the type of waves
emitted after the slit is due to the possibility of a spatial
forcing of the phase by the incoming plane wave. The latter involves
a typical length, the inverse of the vertical number $k_z$, which is
in the present experiment nothing but the wavenumber forced by the
generator. It appears that a criterion for a change in behavior
occurs when the spatial evolution of the phase is small compared to
the temporal one, leading to $k_za\leq\omega T$, i.e.
$k_za\leq2\pi$. In the present case, it leads to
$a\leq\lambda_e=3.6$ cm. The main question remaining is to find a
precise criterion to discriminate when spatial forcing of the phase
occurs or not. This phenomenon of phase forcing might be related to
circular oscillations of a cylinder in a stratified fluid which
leads to two preferential emission (two beams instead of four).
\cite{Gavrilovermanyuk,HurleyHood}

\section{Conclusion}
\label{conclusion}

In this article, we have applied the complex demodulation, also
called Hilbert transform.  This transformation is shown to be very
powerful when adapted to internal waves in two dimensions. The
experimental investigation of attenuation, reflection and
diffraction of internal plane waves generated using a new type of
generator has brought answers to several theoretical assumptions
never confirmed.

The attenuation of internal plane waves is in good agreement with
the linear viscous theory of internal waves. Furthermore, the
results obtained quantify the influence of the wavelength since we
consider monochromatic internal plane waves.

Although the reflection of internal waves is a classic phenomenon,
some theoretical ideas remained assumptions, and by looking for an
hypothetical back-reflected wave we can now confirm that the
back-reflection is not present.

Finally, we study the problem of diffraction of internal waves as it
has not been investigated to our knowledge yet, and we exhibit the
diffraction pattern of internal wave which is atypical due to the
peculiar dispersion relation of internal waves.

\bigskip{\bf Acknowledgments}

We thank Denis Le Tourneau and Marc Moulin for their helps in
preparing the experimental facility. This work has been partially
supported by the 2006 IDAO-CNRS, 2007 LEFE-CNRS program and by
2005-ANR project TOPOGI-3D.

\bibliographystyle{unsrt}

\begin{thebibliography}{10}

\bibitem{Dalziel2000}
S.~B. Dalziel, G.~O. Hughes, and B.~R. Sutherland.
\newblock Whole-field density measurements by `synthetic schlieren'.
\newblock {\em Experiments in Fluids}, 28:322--335, 2000.

\bibitem{Sutherland1999}
B.~R. Sutherland, S.~B. Dalziel, G.~O. Hughes, and P.~F. Linden.
\newblock Visualization and measurement of internal waves by 'synthetic
  schlieren'. part 1. vertically oscillating cylinder.
\newblock {\em Journal of Fluid Mechanics}, 390:93--126, 1999.

\bibitem{PeacockWeidman}
T. Peacock and P. Weidman
\newblock The effect of rotation on conical internal wave beams.
\newblock {\em Expteriments in Fluids}, 39:32 2005.

\bibitem{PeacockTabei}
T. Peacock and A. Tabaei.\newblock Visualization of nonlinear
effects in reflecting internal wave beams. \newblock {\em Physics of
Fluids} 17:061702, 2005.

\bibitem{NyeDalziel}
A. Nye and S. B.  Dalziel. \newblock Scattering of internal gravity
waves from rough topography. \newblock {\em In Proceedings of the
6th International Symposium on Stratified Flows}, Perth. Ed. G.
Ivey. 631-636, 2006.

\bibitem{Gostiaux_these}
L.~Gostiaux.
\newblock {\em \'{E}tude exp\'erimentale des ondes de gravit\'e internes en
  pr\'esence de topographie. \'{E}mission, propagation, r\'eflexion.}
\newblock PhD thesis, ENS Lyon, 2006.

\bibitem{GostiauxDauxois2007}
L.~Gostiaux and T.~Dauxois.
\newblock Laboratory experiments on the
generation of internal tidal beams over steep slopes.
\newblock {\em Physics of Fluids} 19: 028102 (2007).


\bibitem{Peacock2008}
T.~Peacock, P.~Echeverri, and N.~J. Balmforth.
\newblock An experimental investigation of internal waves beam generation by
  two-dimensional topography.
\newblock {\em Journal of Physical Oceanography} 38:235-242, 2008.

\bibitem{K90}
P.~K. Kundu.
\newblock {\em {Fluid Mechanics}}.
\newblock {Academic Press}, 1990.

\bibitem{Croquette:89:2}
V.~Croquette and H.~Williams.
\newblock Nonlinear waves of the oscillatory instability on finite convective
  rolls.
\newblock {\em Physica D}, 37:300--314, 1989.

\bibitem{Garnier:03:1}
N.~B. Garnier, A.~Chiffaudel, F.~Daviaud, and A.~Prigent.
\newblock Nonlinear dynamics of waves and modulated waves in 1d thermocapillary
  flows. i. general presentation and periodic solutions.
\newblock {\em Physica D}, 174:1--29, 2003.

\bibitem{Garnier:03:2}
N.~B. Garnier, A.~Chiffaudel, and F.~Daviaud.
\newblock Nonlinear dynamics of waves and modulated waves in 1d thermocapillary
  flows. ii. convective/absolute transitions.
\newblock {\em Physica D}, 174:30--55, 2003.

\bibitem{Gortler1943}
H.~G\"ortler.
\newblock \"Uber eine schwingungserscheinung in fl\"ussigkeiten mit stabiler
  dichteschichtung.
\newblock {\em Zeitschrift f\"ur Angewandte Mathematik und Mechanik}, 23:65--71,
  1943.

\bibitem{Mowbray1967}
D.~E. Mowbray and B.~S.~H. Rarity.
\newblock A theoretical and experimental investigation of the phase
  configuration of internal waves of small amplitude in a density stratified
  fluid.
\newblock {\em Journal of Fluid Mechanics}, 28:1--16, 1967.

\bibitem{Onu2003}
K.~Onu, M.~R. Flynn, and B.~R. Sutherland.
\newblock Schlieren measurement of axisymmetric internal wave amplitudes.
\newblock {\em Experiments in Fluids}, 35:24--31, 2003.

\bibitem{Gostiaux2007}
L.~Gostiaux, H.~Didelle, S.~Mercier, and T.~Dauxois.
\newblock A novel internal waves generator.
\newblock {\em Experiments in Fluids}, 42:123--130, 2007.

\bibitem{Thomas1972}
N.~H. Thomas and T.~N. Stevenson.
\newblock A similarity solution for viscous internal waves.
\newblock {\em Journal of Fluid Mechanics}, 54:495--506, 1972.

\bibitem{Hurley1997part2}
D.~G. Hurley and G.~Keady.
\newblock The generation of internal waves by vibrating elliptic cylinders.
  part2. approximate viscous solution.
\newblock {\em Journal of Fluid Mechanics}, 351:119--138, 1997.

\bibitem{Sutherland2002}
B.~R. Sutherland and P.~F. Linden.
\newblock Internal wave excitation by a vertically oscillating elliptical
  cylinder.
\newblock {\em Physics of Fluids}, 14:721--731, 2002.

\bibitem{Zhang2007}
H.~P. Zhang, B.~King, and H.~L. Swinney.
\newblock Experimental study of internal gravity waves generated by
  supercritical topography.
\newblock {\em Physics of Fluids}, 19:096602, 2007.

\bibitem{Rieutord2001}
M.~Rieutord, B.~Georgeot, and L.~Valdettaro.
\newblock Inertial waves in a rotating spherical shell: Attractors and
  asymptotic spectrum.
\newblock {\em Journal of Fluid Mechanics}, 435:103--144, 2001.

\bibitem{Hazewinkel2007}
J.~Hazewinkel, P.~van Breevoort, S.~B. Dalziel, and L.~R.M. Maas.
\newblock Observations on the wavenumber spectrum and evoution of an internal
  wave attractor.
\newblock {\em Journal of Fluid Mechanics}, 598:373--382, 2008.

\bibitem{Lighthill}
J.~Lighthill.
\newblock {\em Waves in Fluids}.
\newblock Cambridge Mathematical Library, $3^{rd}$ printing edition, 1978.

\bibitem{MercierGostiauxDauxois}
M. Mercier, L. Gostiaux, and T. Dauxois.
\newblock {\em Internal waves generator}.
\newblock In preparation, 2008.

\bibitem{Baines1971part1}
P.~G. Baines.
\newblock The reflexion of internal/inertial waves from bumpy surfaces.
\newblock {\em Journal of Fluid Mechanics}, 46:273--291, 1971.

\bibitem{Baines1971part2}
P.~G. Baines.
\newblock The reflexion of internal/inertial waves from bumpy surfaces. part 2.
  split reflexion and diffraction.
\newblock {\em Journal of Fluid Mechanics}, 49:113--131, 1971.

\bibitem{Sandstrom}
H. Sandstrom.
\newblock The effect of boundary curvature on reflection of internal
waves.
\newblock {M\'emoires Soci\'et\'e Royale des Sciences de Li\`ege},
6\`eme s\'erie, tome IV, 183-190, 1972.

\bibitem{Dauxois1999}
T.~Dauxois and W.~R. Young.
\newblock Near-critical reflection of internal waves.
\newblock {\em Journal of Fluid Mechanics}, 390:271--295, 1999.

\bibitem{LeoMaas}
L. Maas. Private communication, 2005.

\bibitem{Gavrilovermanyuk}
N. V. Gavrilov, E. V. Ermanyuk.
\newblock Internal waves generated by circular translational motion
of a cylinder in a linearly straified fluid.\newblock{\em Journal of
Applied Mechanics and Technical Physics}  38: 224--227, 1996.

\bibitem{HurleyHood}
D. G. Hurley, M. J. Hood.
\newblock The generation of internal waves by vibrating elliptic
cylinders. Part 3. Angular oscillations and comparison of theiry
with recent experimental observations.
\newblock{\em Journal of Fluid Mechanics} 433, 61--75, 2001.

\end{thebibliography}

\end{document}